\documentclass[12pt]{iopart}
\usepackage{graphicx,color}
\usepackage{amssymb}
\usepackage{enumerate}
\usepackage[utf8]{inputenc}
\usepackage[english]{babel}

\begin{document}

\title[]{\textbf {Analysis of the spectroscopy of a hybrid system
composed of a superconducting
flux qubit and diamond NV$^-$ centers}}

\author{H. Cai,$^{1,2}$ Y. Matsuzaki,$^{1}$ K. Kakuyanagi,$^{1}$
H. Toida,$^{1}$ X. Zhu,$^{3}$
N. Mizuochi$^{4}$ K. Nemoto,$^{5}$ K. Semba$^{6}$ W. J. Munro,$^{1}$   S. Saito,$^{1}$ H. Yamaguchi,$^{1,2}$}
\address{
$^{1}$NTT Basic Research Laboratories, NTT Corporation, Atsugi,
Kanagawa, 243-0198, Japan\\
$^{2}$
Department of Physics, Tohoku University, Sendai, Miyagi 980-8578,
Japan \\
$^{3}$Institute of Physics, Chinese Academy of Sciences,
Beijing, 100190, China
\\
$^{4}$Graduate School of Engineering Science, University of Osaka, 1-3
Machikane-yama, Toyonaka, Osaka, 560-8531,
Japan\\
$^{5}$National Institute of Informatics, 2-1-2 Hitotsubashi, Chiyoda-ku, Tokyo, 101-8430, Japan\\
$^{6}$National Institute of Information and Communications Technology,
4-2-1, Nukuikitamachi, Koganei-city,
Tokyo, 184-8795, Japan.\\
}

\ead{matsuzaki.yuichiro@lab.ntt.co.jp}
\begin{abstract}

A hybrid system that combines the advantages of a superconductor flux
 qubit and an electron spin ensemble in diamond is one of the promising
 devices  to realize quantum information processing. Exploring the properties of
 \textcolor{black}{the superconductor diamond system is essential for the efficient use
 of this device.
 When we perform  spectroscopy of this system,
 significant power
 broadening is observed. 
 However, previous models to describe this system are known to be applicable
only when the power broadening is negligible.}
 Here, we construct a  new
 approach to analyze this system with strong driving, and succeed to reproduce the spectrum
 with the power broadening.
 Our results provide an efficient way to analyze this hybrid system.
   
   
\end{abstract}

\maketitle

\section{Introduction}
The properties of different types of qubits have been studied
extensively over the past decade toward realizing quantum computation.
Each system has its own advantages and disadvantages.  The idea of
combining different systems is pursued by several  groups, which would
inherit the advantages of each system, such as atoms coupled to
optical cavities,  spins coupled to resonators, and spins coupled to superconducting
qubits
\cite{pirkkalainen2013hybrid,rempe2010eit,sorensen2004capacitive,tian2004interfacing,
rabl2006hybrid,marcos2010coupling,laddqubus2006}.  


One of the most successful hybrid schemes is to couple the electron spin
ensemble with superconducting circuit to realize the fast processing and
long storage times.
Although an electron spin ensemble
may possess excellent coherence properties, the weak non-linearity makes
it difficult to realize a large scale quantum computer.  Superconducting
qubits are sensitive to the noise which leads to a relatively short
coherence time. Instead, due to a nonlinearity that Josephson Junctions
has, a superconducting circuit has excellent controllability of quantum
states involved in the processing, such as single qubit gates, two qubit
gates, and projective measurements \cite{
clarke2008superconducting,nakamura1999coherent,vion2002manipulating,chiorescu2003coherent,
sillanpaa2007coherent,majer2007coupling,dicarlo2009demonstration,ansmann2009violation,barends2014superconducting}. There
are two major \textcolor{black}{schemes} that utilize this type of hybrid system. One of
\textcolor{black}{them} consists of a spin ensemble, a superconducting
resonator, and a superconducting transmon qubit
\cite{Imamoglu2009,Wesenberg2009,kubo2011hybrid,kubo2010strong,kubo2012storage}. Here,
the resonator plays the role of a quantum bus \cite{laddqubus2006,majer2007coupling}
where
\textcolor{black}{the resonator mediates the interaction between the spins
and a superconducting qubit.}
The other one is \textcolor{black}{to use} a direct coupling between a spin ensemble and the
superconducting flux qubit (FQ)
\cite{marcos2010coupling,zhu2010coherent,saito2013memory,zhudark2014,matsuzakirabi2015}.

A particularly attractive spin ensemble for such hybrid systems is
negatively charged nitrogen vacancy centers (NV$^-$) in the diamond
crystal whose coherence time
 \textcolor{black}{is around} 0.6 $s$
 \cite{mizuochi2009coherence,balasubramanian2009ultralong}.  An NV$^-$
 center is a defect consisting of a substitutional nitrogen atom and an
 adjacent vacancy \textcolor{black}{where an additional electron is
 trapped. It is known that an NV$^-$ center is a spin-1 system, and}
 \textcolor{black}{there is an energy splitting
 of 2.88 GHz
 between the states $\arrowvert0\rangle$ and $\arrowvert\pm1\rangle$
 without an applied magnetic fields.} Furthermore, the NV$^-$ centers can
 be strongly
 coupled with superconducting circuits by a super-radiant effect\cite{Imamoglu2009,Wesenberg2009}, where
 the coupling strength is
 enhanced by a factor of $\sqrt{N}$.

Several groups have already \textcolor{black}{demonstrated} the hybrid
coupling systems by the ensemble of electron spins and superconducting
circuit resonator
\cite{kubo2011hybrid,kubo2010strong,kubo2012storage,
kubo2012electron,amsuss2011cavity,schuster2010high,putz2014protecting}.
Also,
Further, Kubo et al. \cite{kubo2011hybrid} \textcolor{black}{has
succeeded in indirectly coupling the ensemble to the transmon qubit
using a frequency tunable resonator to mediate it.}
However, in order to enhance the coupling strength, this system requires
 a large number of electron spins $(\sim 10^{12})$ in the ensemble to
 \textcolor{black}{achieve} the coupling strength of tens of MHz \cite{kubo2011hybrid,
 kubo2010strong,kubo2012storage,kubo2012electron,
 amsuss2011cavity,schuster2010high,putz2014protecting}.
 These experiments \textcolor{black}{need} the mm-size resonator, which
 could be difficult for the integration
 of many memories on a single quantum based chip  \cite{kubo2010strong,kubo2012storage}. 

\textcolor{black}{The advantage to use the spin ensemble
and the superconducting flux qubit \cite{marcos2010coupling,zhu2010coherent,saito2013memory}
is that the superconducting flux qubit can couple directly and more
strongly with the spin ensemble}
due to the persistent current of the flux qubit around 300 $nA$ $\sim$
900 $nA$ \cite{paauw2009tuning,zhu2011coherent}.
\textcolor{black}{This system} requires a much smaller number of NV$^-$ centers spins $(\sim
10^{7})$ to realize a strong coupling \textcolor{black}{regime, and the sample occupies only
a surface area of tens of micro-meter square.}
Coherent strong coupling between a flux qubit and nitrogen-vacancy (NV$^-$)
centers has already been demonstrated
\cite{zhu2010coherent,saito2013memory,zhudark2014},
as well as the
coherent exchange of a single quantum excitation \cite{zhu2011coherent}.
Quantum memory operations involving single qubit and entangled states
have been implemented \cite{saito2013memory}.

Some theoretical models have been investigated to describe such hybrid
systems
\cite{marcos2010coupling,Imamoglu2009,Wesenberg2009,diniz2011strongly}.
Traditionally, Jaynes-Cummings model (JC model) is used to describe the
coupling between the flux qubit and the spin ensemble
\cite{marcos2010coupling,Imamoglu2009,Wesenberg2009,diniz2011strongly}. Under
a strong external magnetic field, 
the degenerate state $\arrowvert1\rangle$ and $\arrowvert-1\rangle$ of
the NV$^-$ center would be separated far from each other
\textcolor{black}{and}
the NV$^-$ center can be
considered as a spin $1/2$ system (two-level system).
If the number of excitations in the hybrid system is much smaller than
the number of NV$^-$ centers, we can treat this system as a
harmonic oscillator and then use JC model to describe the coupling
between the flux qubit and NV$^-$ centers.
However, when the applied magnetic field is weak, NV$^-$ center will
show the properties of a spin 1 system (three-level system)
\cite{zhudark2014}
because of \textcolor{black}{the} degenerate states  of $\arrowvert1\rangle$ and
$\arrowvert-1\rangle$. Therefore, the conventional model \textcolor{black}{cannot} be
straight-forwardly applied to analyze the properties of flux qubit coupled with NV$^-$ center for this regime.

From the spectroscopic measurement in a superconductor diamond system
without an applied magnetic field,
three resonant peaks have been observed in \cite{ kubo2010strong,
zhu2011coherent}. A sharp peak
has been clearly observed in the middle of the avoided crossing caused
by the coherent coupling between the superconducting circuits and an
ensemble of electron spins in nitrogen-vacancy center. The center narrow
peak shows a longer lifetime than that of the other two broader peaks,
which may provide an alternative approach for the quantum memory
\cite{zhudark2014,fleischhauer2002quantum}.

\textcolor{black}{In the paper \cite{zhudark2014}}, a full Hamiltonian model for  the hybrid system of the
flux qubit and NV$^-$ centers \textcolor{black}{is} used to interpret the mechanism causing the
sharp narrow peak, which contains the properties of spin 1 system such
as inhomogeneous strain, randomized magnetic field from P1 center
(substitutional nitrogen centers in diamond), and zero field splitting
fluctuation \cite{zhudark2014}. \textcolor{black}{The NV$^-$ center is regarded as a
three-level system, and it was shown} that two collective modes of the NV$^-$
ensembles are relevant for the spectrum. \textcolor{black}{The two broader
peaks are}
associated with the dressed bright state in NV$^-$ that can be directly
coupled with the flux qubit.
The center narrow peak arises from another
specific collective mode that cannot be directly driven by the flux
qubit. \textcolor{black}{This is} called a "dark state". Due to the effects of the strain
and Zeeman splitting, the bright state and dark state can exchange the
excitation, \textcolor{black}{while} the bright states interact with the
flux qubit.
In \textcolor{black}{that} theoretical analysis \cite{zhudark2014},  the
flux qubit \textcolor{black}{is regarded} as a
harmonic oscillator in the limit of low excitation energy so that 
the Heisenberg equations \textcolor{black}{can be solved} in the frequency
domain.
However, this approximation is valid \textcolor{black}{only when}  the driving
power is weak, \textcolor{black}{and} will be invalid to describe the phenomenon when
the power broadening \textcolor{black}{is observed} in the strong driving power case. Since it is
not always possible to obtain a reliable data without power broadening
due to a small signal to noise ratio, this could be a
limitation to describe the hybrid system \textcolor{black}{for such a
regime.}
Therefore, it is necessary to build an
alternative approach that can include the effect of power broadening in
the strong driving regime.

In this paper, we \textcolor{black}{extend the previous approach}
 to reproduce the experimental spectroscopy that is
significantly affected by the power
broadening. \textcolor{black}{The rest of this paper is organized as follows. In section 2,}
\textcolor{black}{we introduce} the previous model for
this hybrid system \cite{zhudark2014}.
\textcolor{black}{In section 3, we
highlight a new theoretical model to show
that the numerical results agree with the experiments. Finally, section 4 contains a summary
of our results.}




\section{Hamiltonian}
\textcolor{black}{Let us review} the
Hamiltonian introduced
by \textcolor{black}{the paper in \cite{zhudark2014}.
The diamond containing NV$^-$ centers is attached on top of the flux qubit so
that these two system can interact each other, as described in the Appendix \ref{experimentsetup}.
The Hamiltonian of this system includes} not only the interaction
between the flux qubit and NV$^-$ centers, but also the
strain
distributions, the inhomogeneous zero field splittings, the effect of $P1$ centers,
and the hyperfine coupling from the nuclear spin, described as follows:
\begin{equation}
  H=H_{\rm{flux}}+H_{\rm{drive}}+H_{\rm{ens}}+H_{\rm{int}}\label{fullhamiltonian}
\end{equation}
The flux qubit Hamiltonian, interaction Hamiltonian, the NV$^-$
diamond ensemble
Hamiltonian and the driving Hamiltonian can be written as:
 \begin{eqnarray}
H_{\rm{flux}}&=&\frac{\hbar}{2}\varepsilon \hat{\sigma }_z +\frac{\hbar}{2}\Delta \hat{\sigma}_x \\
H_{\rm{drive}}&=&\hbar\lambda  \cos \omega t \cdot \hat{\sigma}_z \\
H_{\rm{ens}}&=&\hbar\sum_{k=1}^{N} \Big{ \{ }D_{k}\hat{S}^2_{z,k}+
 E^{(k)}_1(\hat{S}^2_{x,k}-\hat{S}^2_{y,k})+\nonumber \\
 &&E^{(k)}_2(\hat{S}_{x,k}\hat{S}_{y,k}+\hat{S}_{y,k}\hat{S} _{x,k})+
 g_e\mu_B{\bf{B}}^{(k)}_{\rm{NV}} \cdot {\bf{S_k}} \Big{ \} }  \\
H_{\rm{int}}&=&\hbar g_e\mu _B \hat{\sigma }_z
 \Big{(}\sum_{k=1}^{N} {\bf {B}}^{(k)}_{\rm{qb}}\cdot {\bf{S}}_k\Big{)}  
\end{eqnarray}
where $\widehat{\sigma}_{x,z}$ denotes the Pauli matrices for the flux
qubit, $\widehat{\sigma}_{z}$ represents the population difference
between two persistent current states in the flux qubit,  $\vartriangle$
denotes the flux qubits tunneling energy, and $\varepsilon$ denotes the
bias. From the Hamiltonian of the flux qubit,
the 
frequency of the gap tunable flux qubit is calculated as $
\omega_{FQ}= \sqrt{\varepsilon^{2}+\vartriangle^{2}}$. 
Since the bias  $\varepsilon$ can be controlled by the external
magnetic flux, \textcolor{black}{we} use the case of no bias (where
the bias
$\varepsilon$ is equal to 0), in which case the flux qubit has the best
coherence time at the optimal point
\cite{zhu2010coherent,saito2013memory}.  In the expression
 for the driving field of the flux qubit $H_{drive}$, $\lambda$ is the
amplitude of the field, and $\omega$ is the angular frequency of the
microwave. $H_{ens}$ represents the electron spin ensemble of $N$
individual NV$^-$ centers with $\widehat{S}_{x,y,z}$  describing the
electron spin $1$ operators of the individual NV$^-$ center. $H_{ens}$ is
characterized by the zero-field spitting $D$, the strain induced
spitting $E$ and the Zeeman splitting term $g_{e} \mu_{B} B_{NV}\cdot
S$. Here, $g_{e}=2$ is the NV$^-$ Lande factor and $\mu_{B}$ is Bohr's
magneton. $B_{NV}$ is the magnetic field composed by
the inhomogeneous magnetic field from $P1$
centers and hyperfine induced magnetic field generated by nitrogen
nuclear spins. $E^{(k)}_1$ denotes the strain along the x direction
while $E^{(k)}_2$ denote the strain along the y direction.
The term $B_{qb}$ in the $H_{int}$ presents the magnetic
field generated by a flux-qubit persistent current.

\textcolor{black}{Since it is difficult to solve the Hamiltonian described
in the
Eq. \ref{fullhamiltonian},
our previous model 
 considered
the flux qubit and
NV$^-$ centers as harmonic oscillators on the condition that the driving
field strength of the flux qubit is weak \cite{zhudark2014}.
We call this
a many harmonic oscillator
model (MHOM). The details of this model are discussed in the Appendix \ref{mhomappendix}.
}
Moreover, to obtain an intuitive explanation for the spectroscopic
result, \textcolor{black}{the Hamiltonian is further simplified with}
homogeneous NV$^-$ centers under a zero magnetic field $($such as
$D_k=D$, $B_k=B$ and $E_k=E$ $)$ without the driving
term ($\lambda =0$) \cite{zhudark2014}. Considering a
homogeneous system, one can describe the NV ensemble by using just two
harmonic oscillators. The Hamiltonian is rewritten as:
\begin{eqnarray}
H=\hbar \omega _{FQ}\hat{c}^{\dagger }\hat{c}
 +\hbar \omega
 _{b}\hat{b}^{\dagger}\hat{b} +\hbar \omega
 _{d}\hat{d}^{\dagger}\hat{d}
 +\hbar
  g(\hat{c}^{\dagger}\hat{b}+\hat{c}\hat{b}^{\dagger
  })
  +\hbar
  Je^{i\theta }\hat{b}^{\dagger}\hat{d}
  +\hbar Je^{-i\theta }
  \hat{b}\hat{d}^{\dagger }\label{simple}\ \ \ \ \ \ 
\end{eqnarray}
\textcolor{black}{where 
$\omega _b$ ($\omega _d$) denotes the frequency of the bright (dark) mode of the NV$^-$ centers, $\hat{c}^{\dagger }$ denotes a creation
operator of the flux qubit, $\hat{b}^{\dagger }$ ($\hat{d}^{\dagger }$) denotes a creational operator
of the bright (dark) mode of the NV$^-$ centers respectively, $g$ denotes the
coupling strength between the NV$^-$ centers and the flux qubit, $Je^{i\theta} = g_{e} \mu_{B}
B+iE$ denotes the coupling between the bright mode and dark mode. It is worth mentioning
that the bright mode can be coupled with the flux qubit while the dark mode has
no direct coupling with the flux qubit.
The frequency of the bright state is assumed to be equal to that of the
dark state, and so we have $\omega _{NV}=\omega_b=\omega_d=D$.}
 This corresponds to 
the zero-field splitting of the NV$^-$ center between
$\arrowvert0\rangle$ and $\arrowvert\pm1\rangle$ states at zero magnetic
field.
(\textcolor{black}{In the Appendix \ref{eigensimple}, the eigenvalues and
eigenvectors
of this simplified Hamiltonian are
described, and we discuss the
properties of them.})

\section{Main results}

\subsection{Master equation model (ME)}

A spectroscopy of this system was performed, and it was shown that power
broadening due to the strong driving power is relevant in this
system (shown in the \cite{zhudark2014}). There are several models
attempting to understand this system. However, \textcolor{black}{no
existing model can} explain the
power broadening induced by strong microwave pulses for this system
\cite{zhudark2014,zhu2011coherent}.
It is important to build a model to understand
this phenomenon. Such a model would contribute to evaluate the performance of
the hybrid system under different driving powers and would help to optimize
the fabrication parameters.

\begin{figure}[h!]
\par
\centering
\includegraphics[scale=0.3]{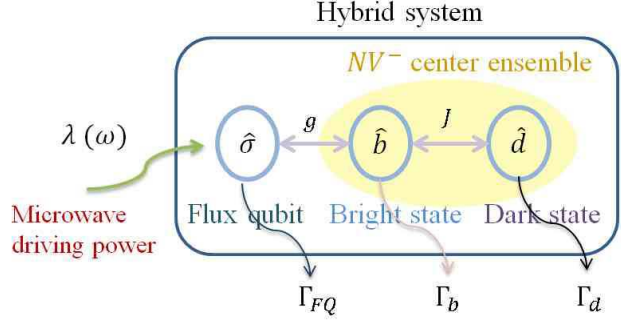}
\par
\caption{Schematic of the model for our approach (ME) to describe
 superconductor diamond hybrid system.
 The flux qubit is considered as a two level system. \textcolor{black}{On the
 other hand, the NV
 centers are replaced by two harmonic oscillators, which we call a
 bright state mode and a dark state mode. The bright
 state is coupled with both the flux qubit and the dark state.}
 }
\label{fig:MEM}
\end{figure}

Here, we introduce our approach using a master equation model (ME)
to study this hybrid system (shown in Fig.\ref{fig:MEM}).
As the ensemble is driven by the flux qubit that is an effective single
photon source,
the number of excitations in the ensemble is much smaller than the
number of electronic spins, and therefore it is
reasonable to regard the spin ensemble as a number of harmonic
oscillators as the previous authors considered.
However, when the external driving power for the flux qubit is strong
and induces the effect of the power broadening,
we should consider the flux qubit as a two-level system and use the spin
$1/2$ operator $\widehat{\sigma}$
instead of the harmonic oscillator operator. This is  because strong
driving power would populate higher energy structure in a harmonic
oscillator system,
which is different from the two-level system.


In such a case, the Hamiltonian should be given without harmonic oscillator approximation:
\begin{eqnarray}
H&=&\frac{1}{2}\hbar (\omega _{FQ}-\omega )\widehat{\sigma}_z+\hbar
 (\omega
 _{NV}-\omega )\hat{b}^{\dagger}\hat{b} +
 \hbar (\omega
 _{NV}-\omega )\hat{d}^{\dagger}\hat{d} +\hbar g
 (\widehat{\sigma}_+\hat{b}+\widehat{\sigma}_-\hat{b}^{\dagger
 })\nonumber \\
&+&\hbar Je^{i\theta} \hat{b}^{\dagger}\hat{d}+\hbar
 Je^{-i\theta}\hat{b}\hat{d}^{\dagger }
 +\frac{\lambda}{2}\widehat{\sigma}_x \label{ourhamiltonian}
\end{eqnarray}
Moreover, we use the dissipative term of Lindblad equation
\cite{ford1999comment,prosen2010spectral,ishizaki2009adequacy} to
present the effect of inhomogeneous broadening induced by the strain and
Zeeman splitting from each NV$^-$ center:
\begin{equation}
 \frac{\partial \rho}{\partial t}=-\frac{i}{\hbar}[\rho,H]+L(\rho)
\end{equation}
\begin{eqnarray}
L(\rho)&=&-\Gamma_{FQ}(\widehat{\sigma}_+\widehat{\sigma}_-\rho+
 \rho\widehat{\sigma}_+\widehat{\sigma}_--2\widehat{\sigma}_-\rho\widehat{\sigma}_+)\nonumber \\
 &-&\Gamma_{b}(\hat{b}^{\dagger}\hat{b}\rho+\rho\hat{b}^{\dagger}\hat{b}-2\hat{b}\rho\hat{b}^{\dagger})
  -\Gamma_{d}(\hat{d}^{\dagger}\hat{d}\rho+\rho\hat{d}^{\dagger}\hat{d}-2\hat{d}\rho\hat{d}^{\dagger})
\end{eqnarray}
where $L(\rho)$ denotes a Lindblad super-operator and $\rho$ is the density
matrix.  $\Gamma_{FQ}$, $\Gamma_{b}$, and $\Gamma_{d}$ denote the decay rate of
the flux qubit, bright state, and dark state, respectively.
\textcolor{black}{The}
inhomogeneous broadening induces unknown phases on the bright state (or
dark state), which transforms the collective mode into another orthogonal
modes \cite{saito2013memory,zhudark2014,zhu2011coherent}.
\textcolor{black}{It has been shown that an inhomogeneous distribution of
the frequency can be considered as 
an energy relaxation of the collective mode \cite{diniz2011strongly}, which can be described as a
Markovian process. Actually,
 an exponential
decay of the collective mode in this hybrid system has been observed in
\cite{saito2013memory}, which is consistent with the model of the
Markovian noise.
  So, to include the effect of inhomogeneous broadening, we add the
  Markovian relaxation terms of the
  bright state and the dark state in the Lindblad equation. }

\subsection{Three Harmonic Oscillator model (THOM)}

In order to describe our sample of the flux qubit coupled with the
NV$^-$ centers,
we have already obtained the necessary parameters $(\zeta,\delta
B,\delta E,\delta D)$
for the simulation of the MHOM from the experiment \cite{zhudark2014}.
However, as we describe before, this model will be invalid for a strong
driving regime,
\textcolor{black}{and we should use the ME when the power broadening is relevant.}
Since we use a different model, we cannot straightforwardly use the
parameters that are used in MHOM.
We need to find suitable parameter set
$(g,J,\Gamma_{FQ},\Gamma_{b},\Gamma_{d})$ for \textcolor{black}{the ME}.
For this purpose, it is convenient to have a fitting function that let
us
find out such a new parameter set
$(g,J,\Gamma_{FQ},\Gamma_{b},\Gamma_{d})$
from the known parameter set $(\zeta,\delta B,\delta E,\delta D)$.
\textcolor{black}{In the condition of weak excitation,
the excited probability of the qubit is described as
\begin{equation}
\langle\hat{\sigma }_+\hat{\sigma }_-\rangle=(\frac{\lambda}{2})^{2}\times
 |\frac{[(i\Gamma_{b}-\omega^{'}_{b})(i\Gamma_{d}-\omega^{'}_{d})-J^{2}]}
 {(i\Gamma_{c}-\omega^{'}_{c})[(i\Gamma_{b}-\omega^{'}_{b})(i\Gamma_{d}-\omega^{'}_{d})-J^{2}]-g^{2}
 (i\Gamma_{d}-\omega^{'}_{d})}|^{2}\label{fittingfunctionthom}
\end{equation}
where $\omega^{'} _{FQ}=\omega_{FQ}-\omega$ and $\omega^{'}
_{NV}=\omega_{NV}-\omega$. We can derive this from a three harmonic
oscillator model (THOM) where the flux qubit is regarded as a harmonic oscillator (See the Appendix \ref{thomappendix} for the details).
It is worth mentioning that
this analytical solution is useful only for weak driving limit. However,
this approach plays a crucial role to find the parameter set for the
master equation approach, as we describe later.
}

\subsection{Determination of parameters for ME}

In this section, we will introduce the process about how to determine
the necessary parameters
to describe our devices for our ME. We have already obtained the
parameters to describe our
devices for the MHOM (the model described in the paper \cite{zhudark2014}):
\begin{equation}
\delta B/2\pi=0.056 \ \rm{mT};\ \delta E/2\pi=4.4 \ \rm{MHz};\ \delta D/2\pi=0.2 \ \rm{MHz};
\end{equation}
\textcolor{black}{We construct a scheme} to translate the parameters $(\zeta,
\delta B, \delta E, \delta D)$ in the MHOM  into the
parameters in our new
approach $(g, J, \Gamma_{b}, \Gamma_{d})$,
\textcolor{black}{which is one of the main results in our paper}.  As mentioned
before, THOM provides the fitting function to obtain such parameters. If
we fit the MHOM spectrum by this function, it seems that we might
determine the parameters needed in the ME because THOM is equivalent to
ME for a weak driving regime. However, due to many parameters to be
estimated, if we naively apply the fitting function to the MHOM, we
cannot obtain a unique set of parameters from the fitting result. In
general, it requires other methods to fix the set of parameters, and we
have developed such method. Firstly, from the $T1$ measurement, we
estimate the value of the flux qubit decay rate $\Gamma_{FQ}$. It is
worth mentioning that $\Gamma _{FQ}$ denotes a decay rate of the flux
qubit for both MHOM and ME so that we could use the same value for
$\Gamma _{FQ}$ in the ME as that used in MHOM
($\Gamma_{FQ}=\frac{1}{2T_1}=0.33$ MHz). Secondly, by using the
expression of eigenvalues and eigenvectors of the Hamiltonian
\textcolor{black}{described in Eq. \ref{ourhamiltonian},}
we find the
method to estimate the values of $J$ and $g$ from MHOM. Finally, by
fixing the values of $J, g$, and $\Gamma_{FQ}$, we use the analytical
solution of THOM to fit the MHOM's result and determine the other two
parameters of  $\Gamma_{b}$  and $\Gamma_{d}$. The detail of the
procedure is described as follows.

\subsubsection{Estimation of the coupling strength $J$ and $g$}
We explain how to determine the coupling strength $J$ and $g$ from the MHOM.


First, \textcolor{black}{by the Hamiltonian in the Eq. \ref{ourhamiltonian}, the frequency difference between the
right peak and left peak on the resonant condition depends on} the coupling strength $J$ and
$g$ ( $\delta=0$ ), as
shown in the Appendix \ref{eigensimple}.
\textcolor{black}{The relationship is described as}
\begin{equation}
(E_{right}-E_{left})/\hbar=2  \sqrt{g^2+J^2}\label{gtwojtwo}
\end{equation}
\textcolor{black}{Also, if we assume $g\gg \Gamma _b, \Gamma _d, \Gamma
_c$ in the Eq. \ref{fittingfunctionthom}, we can calculate
the frequency difference
between the left and right peak
\begin{eqnarray}
 (E_{right}-E_{left})/\hbar\simeq \sqrt{4(g^2+J^2)-(\Gamma _b-\Gamma
  _d)^2}\simeq 2\sqrt{g^2+J^2}
\end{eqnarray}
and this is
consistent with the Eq. \ref{gtwojtwo}.}
Moreover, from the frequency difference
between the left and right peak \textcolor{black}{plotted by MHOM} (shown in Fig. \ref{fig:energy spectrum
MHOM}),
we obtain the following relationship 
\begin{equation}
 (E_{right}-E_{left})/\hbar
 \simeq 27 \ \rm{MHz}
 \Longrightarrow  \sqrt{g^2+J^2}\simeq 13.5\times2\pi \ \rm{MHz}
\label{equ:gj1}
\end{equation}

\begin{figure}[h!]
\par
\centering
\includegraphics[scale=0.28]{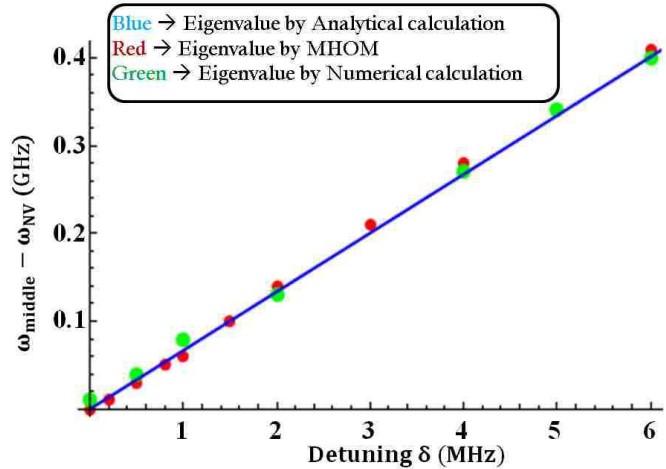}
\par
\caption{Energy shift of the middle peak in the spectroscopy against the
 detuning. Here, the red dots represent the simulation result with the
 MHOM,
 and the green dots represent \textcolor{black}{numerically calculated}
 eigenvalues. The blue line \textcolor{black}{represents}
 an analytical solution using a perturbation theory. }
\label{fig:fitting}
\end{figure}

Second, 
\textcolor{black}{we use}
          a relationship between the middle peak frequency and detuning
          $\delta$.
          \textcolor{black}{We have shown that} the middle peak frequency of the
spectroscopy will be shifted by adding the energy detuning, and
the energy shift increases linearly as we enlarge the detuning
\textcolor{black}{when} the detuning is considered as a perturbation
($\delta\ll\omega_{NV}$) (See the Appendix \ref{thomappendix}).
          This is described as
\begin{equation}
 \omega_{middle}-\omega_{NV}\simeq \frac{\delta J^2}{g^2+J^2}
	 \label{equ:fitting}
\end{equation}
where the proof is described in the Appendix \ref{thomappendix}.
\textcolor{black}{Actually, we plot the energy shift of the middle peak against the detuning by
solving MHOM (shown in Fig.\ref{fig:fitting}), and
we fit this plot by Eq. \ref{equ:fitting}. }
From this fitting, we obtain the following relationship between the parameters for ME:
\begin{equation}
\frac{J^2}{g^2+J^2}\approx 0.067 \ \
 \Longrightarrow  J\simeq 0.27g \label{equ:gj2}
\end{equation}
However, since we use a perturbation to obtain the fitting function
in Eq.\ref{equ:fitting}, this result would be invalid for a
large detuning. To confirm the validity of the perturbation, we
numerically calculate the eigenvalues of the Hamiltonian
\textcolor{black}{in the Eq. \ref{ourhamiltonian}},
 and plot the results \textcolor{black}{in
Fig. \ref{fig:fitting}}.
Since we have a good agreement between perturbation calculation and
numerical results,
we conclude that the perturbation is valid in this parameter regime.

By combining the Eq. \ref{equ:gj1} and Eq. \ref{equ:gj2}, we can estimate the values of J and $g$: 
\begin{equation}
J\simeq 3.5\times 2\pi \ \rm{MHz}; \ \ g\simeq 13\times 2\pi \ \rm{MHz};
\end{equation}

\subsubsection{Estimation of decay rate $\Gamma_b$ and $\Gamma_d$}
The value of $\Gamma_b$ mainly affects the width
of the side peaks while the width of the middle peak is determined by
the value of $\Gamma_d$ (See the Appendix \ref{thomappendix} for the details).
\textcolor{black}{This means that},  we can use the THOM as a fitting function to
fit the spectrum reproduced by the MHOM if the parameters
$\Gamma_{FQ}$, $J$ and $g$ are known,
 and we have obtained the following parameters
\begin{equation}
 \Gamma_{d}\simeq 0.49\times 2\pi \
 \rm{MHz}; \Gamma_{b}\simeq 6.4\times 2\pi
 \ \rm{MHz}; 
\end{equation}
\textcolor{black}{Therefore, all necessary parameters for the ME have been
estimated by our scheme.}

\subsection{Reproducing experimental results with the ME}
\begin{figure}[h!]
\par
\centering
\includegraphics[scale=0.4]{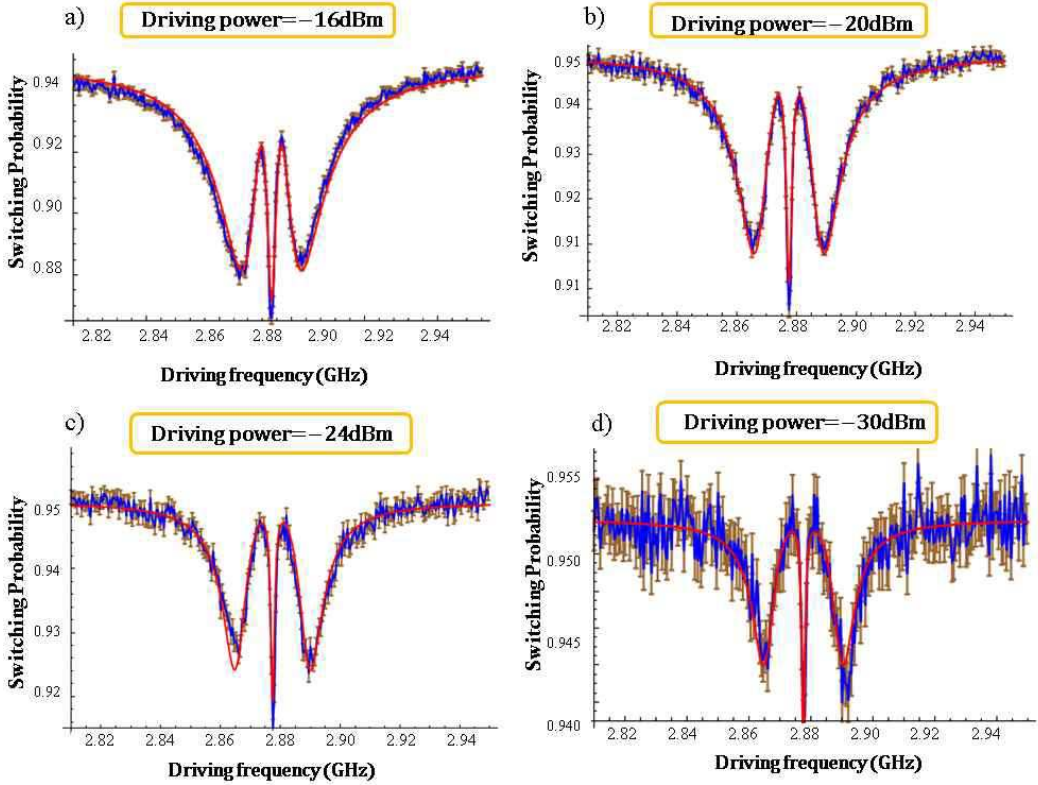}
\par
\caption{Spectroscopy with different driving power. The blue dots are
 the measured data where continuous lines are drawn through the points
 as a guide to the eye.
 The brown line denotes the error bar.
 The red curve shows the numerical results by solving the ME.
 \textcolor{black}{For the simulation, we use parameters of}
 $\omega_{FQ}=\omega_{NV}=2.878\times2\pi$ GHz, $g=12.95\times2\pi$
 MHz, $J=3.46\times2\pi$ MHz,
 $\Gamma_{FQ}=0.300\times2\pi$ MHz, $\Gamma_{d}=0.493\times2\pi$ MHz, and
 $\Gamma_{b}=6.433\times2\pi$ MHz.
 Here, $a$, $b$, $c$, and $d$ present the spectrum in different
 driving power:
 $-16$ dBm, $-20$ dBm, $-24$ dBm and $-30$ dBm, respectively.
 \textcolor{black}{It is worth mentioning that we cannot observe a power
 broadening around -30dBm as described in Fig. \ref{fig:midpeak}, and so
 the power is sufficiently weak in this regime.}
 }
\label{fig:experiement}
\end{figure}
\begin{figure}[ht!]
\par
\centering
\includegraphics[scale=0.42]{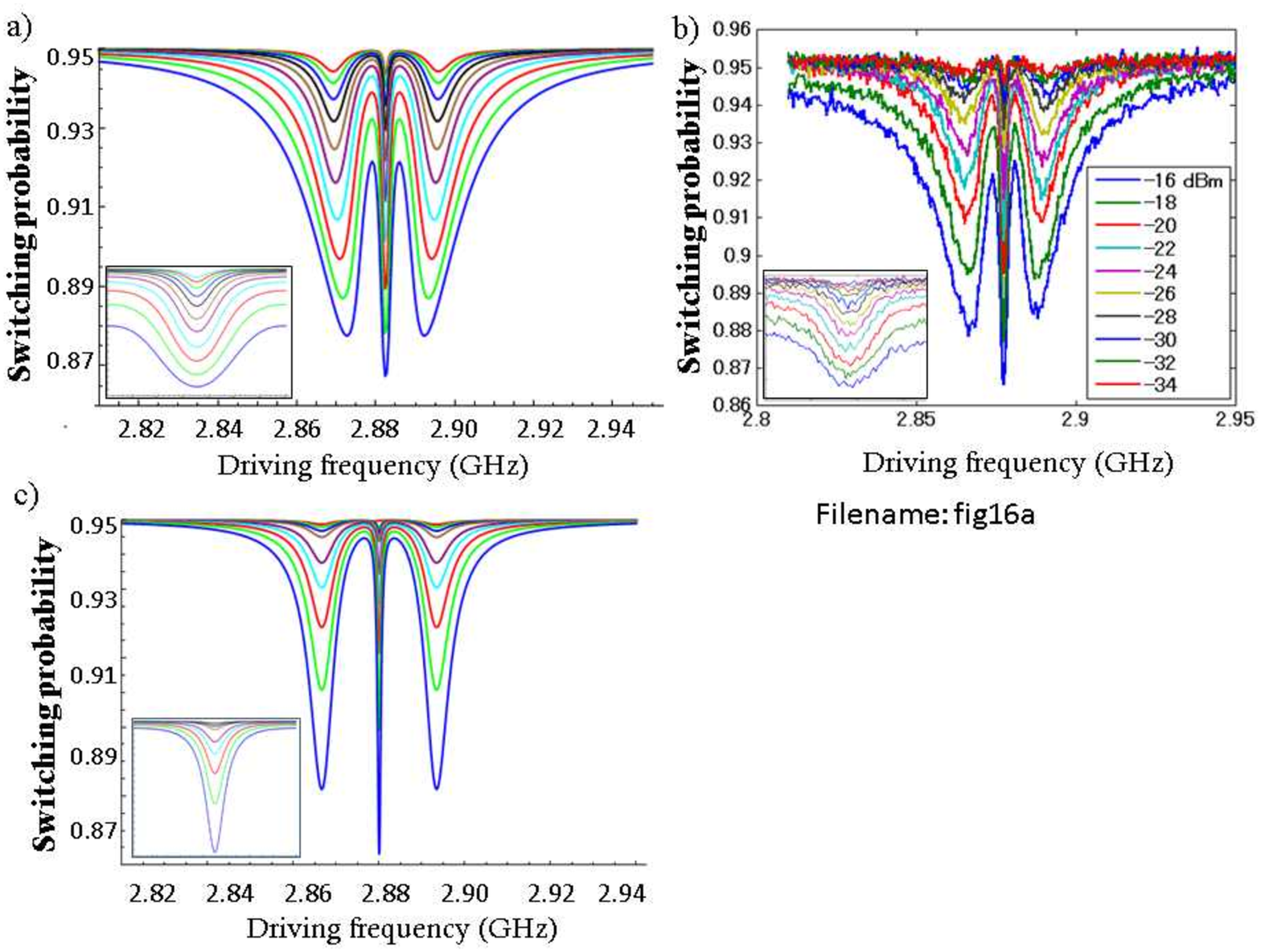}
\par
\caption{The spectrum with power broadening. (a) Power dependence of the
 energy spectrum of our hybrid system obtained from the experiment. (b)
 Numerical results of the energy spectrum by solving ME.
 (c) Numerical results of the energy spectrum by solving MHOM.
 We use the same parameters as those in Fig.\ref{fig:experiement}.
 Inset of each picture is the enlargement of the spectrum over the middle peak region.}
\label{fig:3D}
\end{figure}

With these parameters, we use the ME to reproduce the spectroscopic
measurement in experiment for \textcolor{black}{different} driving power (See
Fig.\ref{fig:experiement} and Fig.\ref{fig:3D}).
Not
only in the weak power case but also in the strong driving power case,
\textcolor{black}{there is a good agreement between
simulation and experiment, as
shown in Fig.\ref{fig:experiement} and Fig.\ref{fig:3D}.}
Thus, our new approach with the ME is shown to be
useful to reproduce the spectroscopy even when the power broadening is
relevant.
\begin{figure}[ht!]
\par
\centering
\includegraphics[scale=0.42]{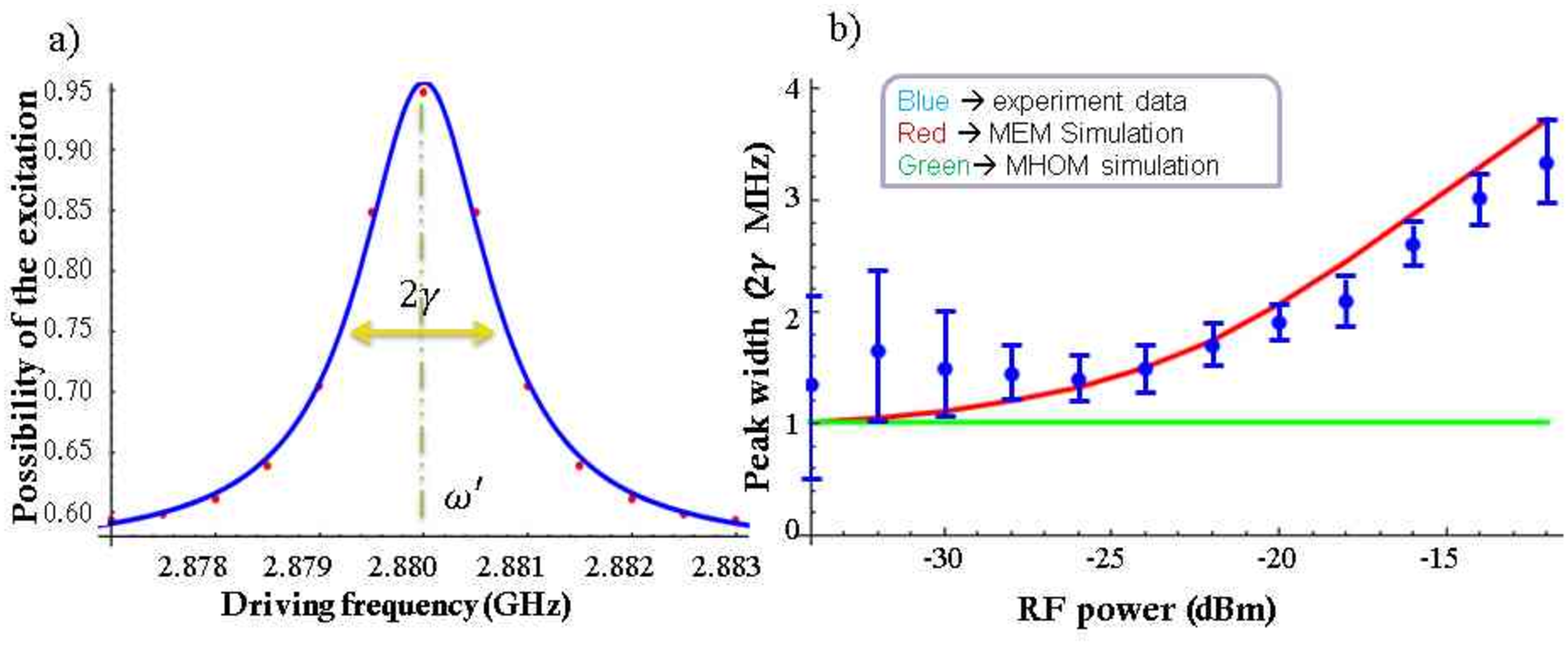}
\par
\caption{Behavior of the width of the middle peak with different driving
 power. (a)
 The middle peak is fit by the Lorentz function. The red dots denote the
 simulation result from ME.
 The blue curve denotes the fitting result of a Lorentz function. The
 parameters for ME are the same as Fig 16.
 (b) The full width at half maximum (FWHM) with different driving power.
 The blue dots present the FWHM from the experiment with error bar.
 The red line denotes the FWHM from the ME by Lorentz fitting.
 The green line is the FWHM resulted from MHOM simulation.}
\label{fig:midpeak}
\end{figure}

Furthermore, in order to compare
the theory and experiment in detail, we analyze the width of the middle
peak which can be well fitted by a
Lorentz function. 
\begin{equation}
 \frac{a\gamma^2}{(\omega-\omega^{'})^{2}+\gamma^2}+c
\end{equation}
where $\gamma$ correspond to the width of the Lorentz curve and
$\omega^{'}$ is the center frequency, $a$ and $c$ and the fitting coefficients.
\textcolor{black}{We use
the Lorentzian function to fit the middle peak
calculated by the ME} (shown in Fig.\ref{fig:midpeak} a). By the
fitting, we obtain the FWHM (Full Width at Half Maximum)
$2\gamma$ \textcolor{black}{with the ME, MHOM, and the experiments, respectively.}
In the  Fig.\ref{fig:midpeak}(b), \textcolor{black}{we plot them, and
the power broadening is clearly observed in both the experiment and the
ME,
while MHOM cannot describe the power broadening. These results also show
that the ME surpasses the MHOM in order to reproduce the experiments.}

\section{Conclusion}
In summary, we introduce a new theoretical approach to describe
spectroscopic measurements of a superconducting flux qubit coupled with
NV$^-$ centers in diamond. Although
\textcolor{black}{previous models are applicable}
only when the driving power of the applied microwave is weak, we
have succeeded to reproduce the experimental spectroscopy even when the
power broadening becomes relevant due to the strong driving. Since it is
typically difficult to remove the effect of the power broadening in the
spectroscopy when a superconducting flux qubit is driven by the
microwave, our results provide an efficient way to analyze the
superconductor diamond hybrid system.
\textcolor{black}{Our method} will be useful to characterize this system for the application of quantum information processing.

\section*{Acknowledgments}

This work was supported by KAKENHI(S) 25220601. The research results
have been achieved by the Commissioned Research of National Institute of
Information and Communications Technology (NICT), JAPAN.

\section{Appendix}

\subsection{Experimental setup}\label{experimentsetup}
\begin{figure}[ht!]
\par
\centering
\includegraphics[scale=0.42]{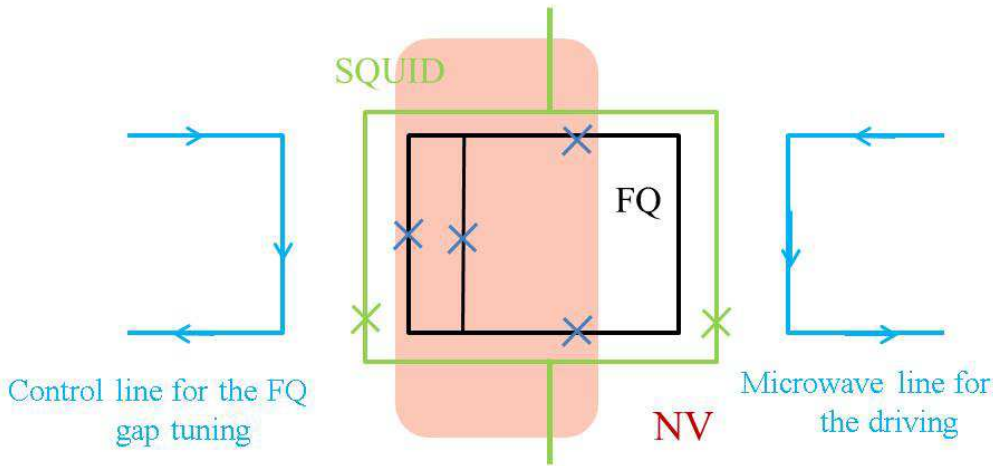}
\par
\caption{Illustration of the hybrid system. It contains the SQUID, the
 control line, microwave line, a flux qubit with four junctions (the
 cross in the diagram),  and an ensemble of NV center in the diamond
 that
 is glued on the top of the flux qubit \cite{zhu2011coherent}.}
\label{fig:MHOM}
\end{figure}

\textcolor{black}{We briefly describe the experimental setup.
The hybrid system is shown in Fig.\ref{fig:MHOM}.
The NV$^-$ ensemble is generated by an ion implantation and annealing in
vacuum \cite{zhu2011coherent}. The density of the NV$^-$ centers is approximately
$5\times 10^{17}$ cm$^{-3}$.
The diamond including the NV$^-$ centers is glued on
the top of the flux qubit where the distance
between the flux qubit and the surface of the NV$^-$ diamond is less
than 1 $\mu $m. We can control the hybrid system by the
microwave, and this system is  measured by switching current in the SQUID that is
inductively coupled with the flux qubit.}

\subsection{Many Harmonic Oscillator model (MHOM)}\label{mhomappendix}
There are two groups that
performed spectroscopy of a superconductor-diamond hybrid system under
zero external magnetic field
\cite{kubo2010strong,zhudark2014,zhu2011coherent}, and both groups
observed two broader side peaks and a middle sharp peak in
\textcolor{black}{the} spectrum. However, in the papers
\cite{kubo2010strong,zhu2011coherent}, \textcolor{black}{such three peaks}
could not be reproduced by the theoretical
model. \textcolor{black}{In a paper \cite{zhudark2014}, 
such three peaks are firstly reproduced by a numerical simulation based
on a new model.}

\begin{figure}[ht!]
\par
\centering
\includegraphics[scale=0.25]{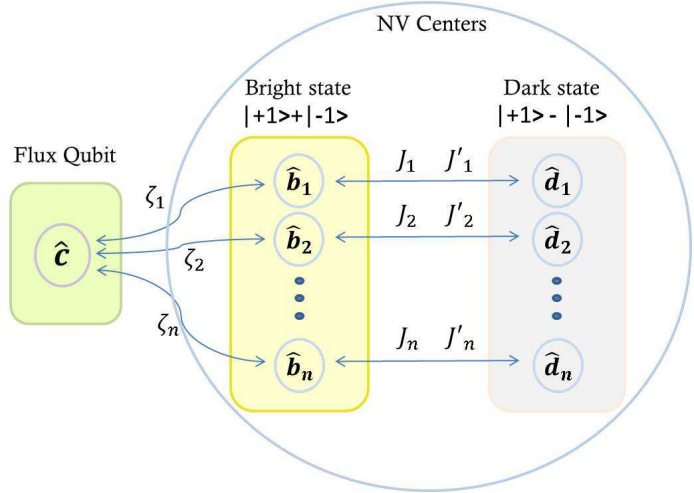}
\par
\caption{The schematic of the MHOM. Here, $\hat{c}$ denotes
 the flux qubit, ${\omega^{(k)}}_b$ (${\omega^{(k)}}_d$)
 denotes
 the frequency of the $k$th bright state (a dark state) in NV$^-$ centers,
 $\hat{b}_k^{\dagger}$ ($\hat{d}_k^{\dagger}$) denotes the bright
 state (dark state) of the $k$th NV$^-$ centers,
 $J_{k}$ denotes the effect from Zeeman splitting, $J^{'}_{k}$
 denotes the effect from the strain, and $\zeta_{k}$ denotes
 the coupling between the flux qubit and $k$th NV$^-$ centers.}
\label{fig:MHOM2}
\end{figure}

\textcolor{black}{We explain how to obtain the model introduced in \cite{zhudark2014}.}
In order to solve the full Hamiltonian described in Eq. \ref{fullhamiltonian},  the flux qubit and
NV$^-$ centers \textcolor{black}{are regarded} as harmonic oscillators on the condition that the driving
field strength of the flux qubit is weak \cite{zhudark2014}. In this
case, the average number of the excitation at the flux qubit and NV$^-$
center is much less than one. Since there are two types of excited
states in the NV$^-$ center ensemble, two creation operators
(such as $\widehat{b}^{+}$  and $\widehat{d}^{+}$ ) \textcolor{black}{are defined} to describe the
properties of NV$^-$ centers (shown in Fig.\ref{fig:MHOM2}). Here,
$\widehat{b}^{+}$ denotes a creation operator of the bright states that
can be directly coupled with the flux qubit, while $\widehat{d}^{+}$
denotes a creation operator of the dark states that interact only
indirectly with the flux qubit via the bright states.  Then making the
rotating wave approximation, \textcolor{black}{we} define the many harmonic oscillator
model (MHOM) and obtain the Hamiltonian as following:
\begin{eqnarray}
H=\hbar (\omega _{FQ}-\omega)\hat{c}^{\dagger }\hat{c}+\hbar
 \frac{\lambda }{2}(\hat{c}+\hat{c}^{\dagger })\nonumber \\
 +\sum_{k=1}^{N}
 \Big[ \hbar (\omega _{b}^{(k)}-\omega)\hat{b}_k^{\dagger}\hat{b}_k
 +\hbar (\omega _{d}^{(k)}-\omega)\hat{d}_k^{\dagger}\hat{d}_k +\hbar
 \zeta_{k} (\hat{c}^{\dagger}\hat{b}_k+\hat{c}\hat{b}^{\dagger
 }_k)\nonumber \\
 +\hbar (J_{k}+iJ^{'}_{k}) \hat{b}^{\dagger}_k\hat{d}_k+\hbar
  (J_{k}-iJ^{'}_{k})\hat{b}_k\hat{d}^{\dagger }_k\Big ]\label{mhom}
\end{eqnarray}
where \textcolor{black}{
${\omega^{(k)}}_b= D_k-E^{(k)}_{1}$ denotes a frequency of the $k$th bright
state and
${\omega^{(k)}}_d= D_k+E^{(k)}_{1}$ denotes a frequency of the $k$th dark
state.
}
Since the effect of the
the Zeeman splitting (strain)
allows the transition between the bright
states and dark states, \textcolor{black}{a coupling strength
of $J_{k}=g_{e} \mu_{B} B_{k}$ ($J^{'}_{k}=E^{(k)}_{2}$) is defined to present
the interaction between them.
} Similarly,
\textcolor{black}{the two-level system of the flux qubit is simplified} as
a harmonic oscillator, and so this can be described by the creational
operator $\widehat{c}^{+}$.  The flux qubit is coupled only with one
mode of the ensemble (the bright states) and the individual coupling
strength is presented by the value of $\zeta$.  $\lambda$ denotes the
driving power of the flux qubit while $\omega$ denotes the  frequency of
the driving power.

\subsection{Heisenberg-Langevin equations of MHOM}\label{mhomappendix}
We explain the details about how we can solve the model introduced in
\cite{zhudark2014}, which we call MHOM.
Based on the Hamiltonian in Eq. \ref{mhom}, the  Heisenberg-Langevin equations of the system are written as \cite{zhudark2014}:
\begin{eqnarray}
\frac{d}{dt}\widehat{c}&=& -(\Gamma_{FQ}+i\omega_{FQ})\hat{c}-i\Big{(}\sum_{k=1}^{N}\zeta_{k}\cdot \hat{b}_k \Big{)}-i\frac{\lambda}{2} \\
\frac{d}{dt}\widehat{b}_k&=& -(\Gamma_b+i\omega_{b}^{(k)})\hat{b}_k-(iJ_k+J^{'}_k)\hat{d}_k-i\zeta_k\cdot \hat{c} \\
\frac{d}{dt}\widehat{d}_k&=& -(\Gamma_d+ i\omega_{d}^{(k)})\hat{d}_k-(iJ_k+J^{'}_k)\hat{b}_k
\end{eqnarray}
where $\Gamma_{FQ}$, $\Gamma_{b}$, $\Gamma_{d}$ denote the energy decay
of the flux qubit and the NV$^-$ center bright state and dark state,
respectively. By transforming these into Fourier space, then \textcolor{black}{we}
obtained the expression of $\hat{c}$:
\begin{equation}
\hat{c}=\frac{\lambda}{2}\times\frac{1}{\omega-\omega_{FQ}+i\Gamma_{FQ}-\Big{(}\sum_{k=1}^{N}\arrowvert\zeta_{k}\arrowvert^{2}
 \frac{\omega-\omega_{d}^{(k)}+i\Gamma_{d}}{(\omega-\omega_{b}^{(k)}+i\Gamma_{b})(\omega-\omega_{d}^{(k)}+i\Gamma_{d})-(J_{k}^{2}+{J^{'}}_{k}^{2})} \Big{)}}
\end{equation}

\textcolor{black}{These Heisenberg-Langevin equations was
solved to obtain $\langle\hat{c}^{\dagger }\hat{c}\rangle$, which denotes the average number
of the excitation in the flux qubit \cite{zhudark2014}.
It is worth
mentioning that the main decoherence source of this hybrid system is
inhomogeneous broadening
\cite{kubo2010strong,kubo2012storage,
zhu2010coherent,saito2013memory,kubo2011hybrid,
kubo2012electron,amsuss2011cavity,schuster2010high,putz2014protecting},
and the Heisenberg-Langevin equations used in \cite{zhudark2014} include
such effect.}
 \textcolor{black}{In this model, the spectroscopic measurements of the flux qubit with
NV$^-$ centers are reproduced} with a narrow peak located in the middle
of the avoid crossing (shown in Fig. \ref{fig:energy spectrum
MHOM}).
\textcolor{black}{Two broad resonances are observed on the side, which
we call a left peak and a right peak, respectively. Also, one sharp resonance is observed
between them around 2.88 GHz, which we call a middle peak. These agree
with the experimental results in the limit of weak driving \cite{kubo2010strong,zhudark2014,zhu2011coherent}.}
\begin{figure}[ht!]
\par
\centering
\includegraphics[scale=0.3]{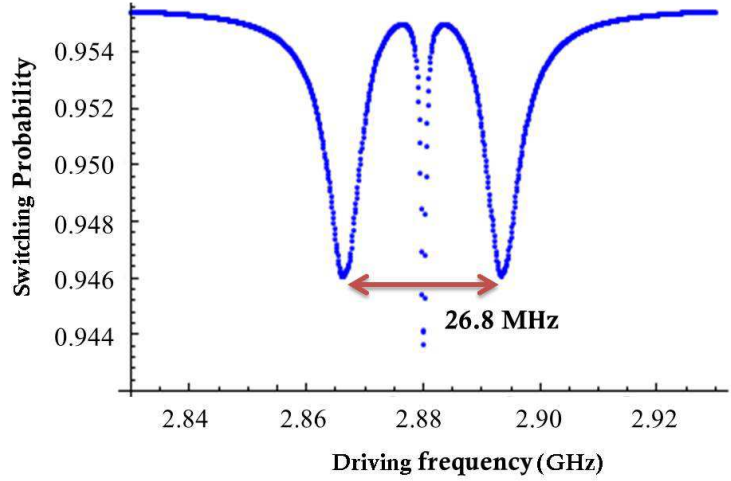}
\par
\caption{The energy spectrum of the hybrid system by MHOM in
 \cite{zhudark2014}. Here, x axis denotes
 the microwave driving frequency and y axis denotes the switching probability
 of the SQUID
 which corresponds to the microwave absorption of the flux qubit
 ($P_{switching}=1-P_{absorption}$). The spectrum is plotted under
 the conditions of $g_{e} \mu_{B} B/2\pi=28$ MHz; $N=36000$;
 $\omega_{FQ}/2\pi=\omega_{b}/2\pi=\omega_{d}/2\pi=2.88$
 GHz; $\delta(g_{e} \mu_{B} B/2\pi)=3.1$ MHz(FWHM); $\delta(E/2\pi)=4.4$
 MHz(FWHM); $\Gamma_{FQ}/2\pi=0.3$
 MHz(FWHM); $\Gamma_{NV}/2\pi=\delta(D/2\pi)=0.2$ MHz(FWHM); $\lambda/2\pi=20$ MHz.}
\label{fig:energy spectrum MHOM}
\end{figure}

\subsection{Eigenvector and Eigenvalue}\label{eigensimple}
\textcolor{black}{To understand the spectrum shown in Fig. \ref{fig:energy spectrum MHOM},
 the Hamiltonian in Eq. \ref{simple} is diagonalised 
 in \cite{zhudark2014}, and we explain the results here.}
\begin{equation}
H=\hbar \omega _{FQ}\hat{c}^{\dagger }\hat{c}+\hbar \omega
 _{NV}\hat{b}^{\dagger}\hat{b} +\hbar \omega
 _{NV}\hat{d}^{\dagger}\hat{d} +\hbar g
 (\hat{c}^{\dagger}\hat{b}+\hat{c}\hat{b}^{\dagger })+\hbar
 J(e^{i\theta}\hat{b}^{\dagger}\hat{d}+e^{-i\theta}\hat{b}\hat{d}^{\dagger})\
 \ \ \ \ \ \ \  
\end{equation}
\textcolor{black}{where} $g$ denotes the collective coupling strength between the bright state
and the flux qubit, J denotes the coupling strength between the bright
state and the dark state,
\begin{figure}[ht!]
\par
\centering
\includegraphics[scale=0.45]{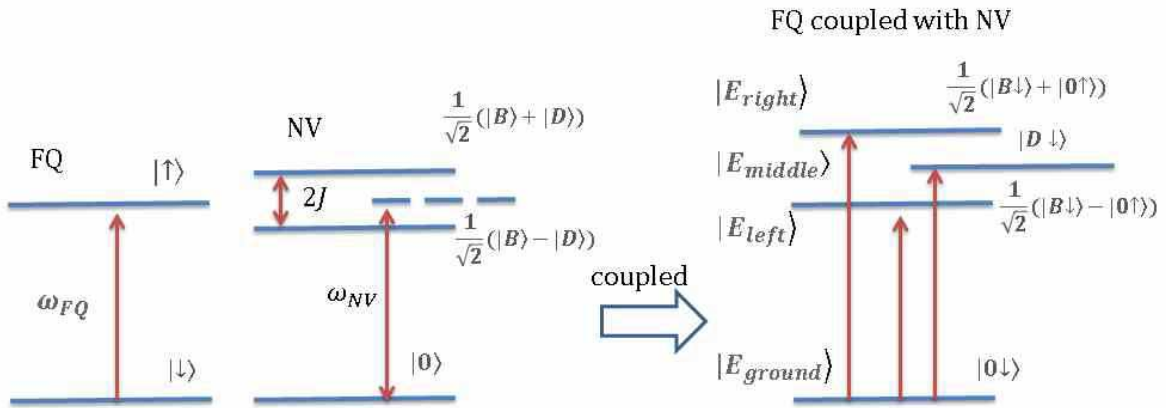}
\par
\caption{The energy diagram. In the picture, the left side
 denotes the energy diagram without coupling (coupling strength $
 g=0 $) while the right side denotes the case with coupling. Here we
 assume that the coupling strength between the dark state and bright
 state
 is much smaller than the coupling strength between the flux qubit and NV$^-$ center $(J\ll g) $.}
\label{fig:energylevel}
\end{figure}
 and $\arrowvert0\rangle$ denotes the ground state of NV$^-$ center.  Since
NV$^-$ center is a spin $1$ system, it has three electron states:
$\arrowvert0\rangle$, $\arrowvert1\rangle$ and
$\arrowvert-1\rangle$.
Suppose that
$\arrowvert B \rangle$ and $\arrowvert D \rangle$ denote the excited
states of bright state and dark state respectively where
\begin{equation}
\arrowvert B \rangle=\frac{1}{\sqrt{2}}(\arrowvert1\rangle+\arrowvert-1\rangle)\end{equation}
\begin{equation}
\arrowvert D \rangle=\frac{1}{\sqrt{2}}(\arrowvert1\rangle-\arrowvert-1\rangle).
\end{equation}
$|\downarrow\rangle$ and $\arrowvert\uparrow\rangle$ denote the ground
state and excited state of the flux qubit.
The relationship between the ground state and the excited state is described as follows:
\begin{equation}
\hat{b}^{\dagger }\arrowvert 0 \rangle_{NV}=\arrowvert B \rangle_{NV}
\end{equation}
\begin{equation}
\hat{d}^{\dagger }\arrowvert 0 \rangle_{NV}=\arrowvert D \rangle_{NV}
\end{equation}
\begin{equation}
\hat{c}^{\dagger }\arrowvert \downarrow \rangle_{FQ}=\arrowvert \uparrow \rangle_{FQ};
\end{equation}
The transition from the ground state to the first excited state (shown
in Fig.\ref{fig:energylevel}),
which corresponds to the left side peak in the experimental spectroscopy (shown in Fig.\ref{fig:energy spectrum MHOM}), is described by:
\begin{equation}
 \bigtriangleup E_{left}=E_{left}-E_0=\hbar(\omega_{NV}-\sqrt{g^{2}+J^{2}})
 \label{equ:nodetuningleft}
\end{equation}
\begin{equation}
 \arrowvert E_{left}\rangle= (-\frac{1}{\sqrt{2}}\arrowvert B
  \downarrow\rangle+\frac{1}{\sqrt{2}}\frac{g}{\sqrt{g^{2}+J^{2}}}\arrowvert
  0
  \uparrow\rangle)+\frac{1}{\sqrt{2}}\frac{Je^{-i\theta}}{\sqrt{g^2+J^2}}\arrowvert D \downarrow\rangle
 \label{equ:leftvector}
\end{equation}
The transition from the ground state to the second excited state (shown
in Fig.\ref{fig:energylevel}), which corresponds to the middle peak
of the spectroscopy (shown in Fig.\ref{fig:energy spectrum MHOM}), is described by:
\begin{equation}
 \bigtriangleup E_{middle}=E_{middle}-E_0=\hbar\omega_{NV}
\end{equation}
\begin{equation}
 \arrowvert E_{middle}\rangle= -\frac{Je^{i\theta}}{\sqrt{g^2+J^2}}\arrowvert 0 \uparrow\rangle+\frac{g}{\sqrt{g^2+J^2}}\arrowvert D \downarrow\rangle
 \end{equation}
 The transition from the ground state to the third excited state (shown in
 Fig.\ref{fig:energylevel}), which corresponds to the right side peak of
 the experimental spectroscopy (shown in Fig.\ref{fig:energy spectrum
 MHOM}), is described by:
 \begin{equation}
 \bigtriangleup E_{right}=E_{right}-E_0=\hbar(\omega_{NV}+\sqrt{g^{2}+J^{2}})
 \label{equ:nodetuningright}
\end{equation}
\begin{equation}
 \arrowvert E_{right}\rangle= (\frac{1}{\sqrt{2}}\arrowvert B
  \downarrow\rangle+\frac{1}{\sqrt{2}}\frac{g}{\sqrt{g^{2}+J^{2}}}
  \arrowvert 0 \uparrow\rangle)+\frac{1}{\sqrt{2}}\frac{Je^{-i\theta}}{\sqrt{g^2+J^2}}\arrowvert D \downarrow\rangle
 \label{equ:rightvector}
\end{equation}

\textcolor{black}{
Due to the hybridization induced by the coupling between the flux qubit
and NV$^-$ centers, the energy level of the exited states of the hybrid
system is split into three levels.}
The left and right peaks in the
Fig.(\ref{fig:energy spectrum MHOM}) correspond to the first and third
excited states which contain a bright state \textcolor{black}{of the NV$^-$ centers}. We
call these two exited states as the hybrid bright states. The middle
peak correspond to the second excited state that shows the existence of
a NV$^-$ dark state, which we call the hybrid dark state. Interestingly,
the second excited state contains $\arrowvert 0 \uparrow\rangle$ where
the flux qubit is excited so that the signal of this excited state can
be detected via the spectroscopic measurement of the flux qubit.

\subsection{Properties of THOM}\label{thomappendix}
We explain how to solve THOM, and discuss the properties of this model.

\subsubsection{Solution of THOM}
\begin{figure}[ht!]
\par
\centering
\includegraphics[scale=0.3]{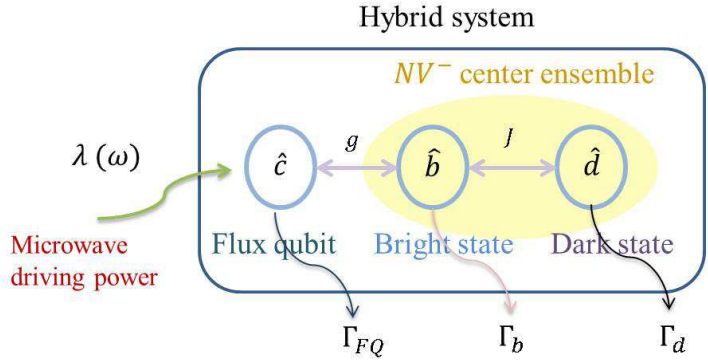}
\par
\caption{Schematic of the model for the THOM to describe superconductor
 diamond hybrid system. The flux qubit is considered as a harmonic
 operator. \textcolor{black}{The bright state is coupled with both the
 flux qubit and
 the dark state.}
 }
\label{fig:THOM}
\end{figure}

In the condition of weak excitation, we can use a harmonic oscillator to
replace the spin $1/2$ operator
$(\widehat{\sigma}_+\Rightarrow\hat{c}^{\dagger})$ and use two harmonic
oscillators $\hat{b}^{\dagger}$ and $\hat{d}^{\dagger}$ to represent
NV$^-$ center. Then, we use the non-hermitian decay terms $(
i\Gamma_{c}\hat{c}^{\dagger}\hat{c}$,
$i\Gamma_{b}\hat{b}^{\dagger}\hat{b}$,
$i\Gamma_{d}\hat{d}^{\dagger}\hat{d}$ ) to take the role of Lindblad
terms, presenting the relaxation in the flux qubit, bright state and
dark state (shown in Fig.\ref{fig:THOM}). 
We rewrite the Hamiltonian as follows.
\begin{eqnarray}
H=\hbar \omega _{FQ}\hat{c}^{\dagger }\hat{c}+
 \hbar \omega_{NV} \hat{b}^{\dagger}\hat{b} +\hbar
 \omega_{NV} \hat{d}^{\dagger}\hat{d} +\hbar g
 (\hat{c}^{\dagger}\hat{b}+\hat{c}\hat{b}^{\dagger })\nonumber \\
 +\hbar
 J(e^{i\theta}\hat{b}^{\dagger}\hat{d}+e^{-i\theta}\hat{b}\hat{d}^{\dagger})
 +\lambda(\hat{c}^{\dagger }+\hat{c})\cos (\omega t)-i\Gamma_{c}\hat{c}^{\dagger}\hat{c}-
  i\Gamma_{b}\hat{b}^{\dagger}\hat{b}-
  i\Gamma_{d}\hat{d}^{\dagger}\hat{d}
  \label{thomhamiltonian}
\end{eqnarray}
, \textcolor{black}{which we call a three harmonic oscillator
model (THOM).}
Here, $\lambda$ denotes the amplitude of the driving power and $\omega$
denotes the frequency of the driving power. $\Gamma$ denotes the energy
relaxation term for flux qubit, bright state and  dark state.


After performing the rotating wave approximation \textcolor{black}{with the Hamiltonian
described in Eq. (\ref{thomhamiltonian})}, we have:
\begin{eqnarray}
H=\hbar \omega^{'} _{FQ}\hat{c}^{\dagger }\hat{c}+\hbar \omega^{'}
 _{NV}\hat{b}^{\dagger}\hat{b} +\hbar \omega^{'}
 _{NV}\hat{d}^{\dagger}\hat{d} +\hbar g
 (\hat{c}^{\dagger}\hat{b}+\hat{c}\hat{b}^{\dagger })+\hbar
 J(e^{i\theta}\hat{b}^{\dagger}\hat{d}+e^{-i\theta}\hat{b}\hat{d}^{\dagger})\nonumber \\
 +\frac{1}{2}\lambda(\hat{c}^{\dagger
  }+\hat{c})-i\Gamma_{c}\hat{c}^{\dagger}\hat{c}-
  i\Gamma_{b}\hat{b}^{\dagger}\hat{b}- i\Gamma_{d}\hat{d}^{\dagger}\hat{d}
\end{eqnarray}
where $\omega^{'} _{FQ}=\omega_{FQ}-\omega$ denotes the difference
between the flux qubit
frequency and driving power frequency. $\omega^{'}
_{NV}=\omega_{NV}-\omega$ denotes
the frequency difference between NV$^-$ center and microwave driving power.

\begin{figure}[ht!]
\par
\centering
\includegraphics[scale=0.4]{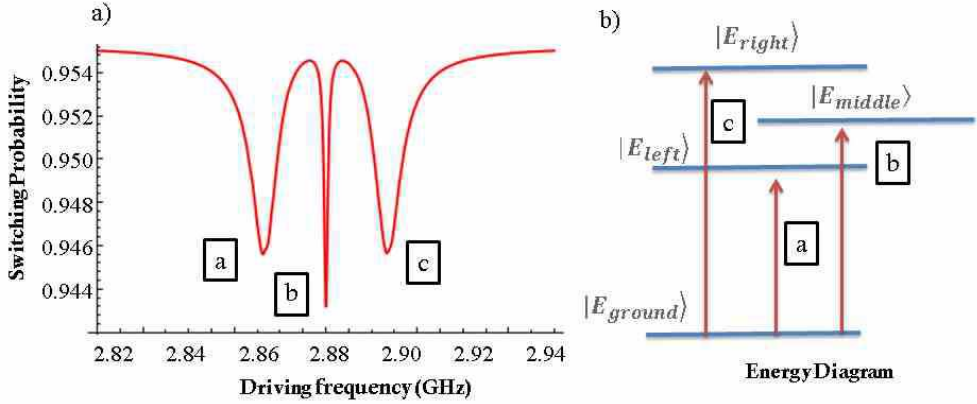}
\par
\caption{The relationship between the spectrum and energy level. In the
 picture (a), we plot the spectrum using the THOM
 \textcolor{black}{where} $\omega_{FQ}/2\pi=\omega_{NV}/2\pi=2.88$ GHz,
 $g=13.0\times2\pi$ MHz, $J=3.46\times2\pi$ MHz,
 $\Gamma_{FQ}=0.30\times2\pi$ MHz, $\Gamma_d=0.50\times2\pi$ MHz,
 $\Gamma_b=6.40\times2\pi$ MHz,  and $\lambda=1.00$ MHz.
 The spectrum denotes the transition from a ground state to each excited
 state.
 In the picture (b), we describe the energy levels diagram.
 \textcolor{black}{The transition between the ground state and the first (third) excited state
 $\arrowvert E_{left}\rangle$ ($\arrowvert E_{right}\rangle$)
 corresponds to the left (right) broad peak in the spectrum.}
 The transition between the ground and the second excited state $\arrowvert
 E_{middle}\rangle$
corresponds to the middle narrow peak.}
\label{fig:THOMspectrum}
\end{figure}

Based on this Hamiltonian, we obtain the following Heisenberg equation:
\begin{eqnarray}
\frac{d}{dt}\widehat{c}&=& -i[\hat{c},H] =  -i(\omega^{'}_{FQ}\hat{c}+g\hat{b}+\lambda-i\Gamma_{c}\hat{c}) \\
\frac{d}{dt}\widehat{b}&=& -i[\hat{b},H] =  -i(\omega^{'}_{NV}\hat{d}+g\hat{c}+Je^{i\theta}\hat{d}-i\Gamma_{b}\hat{b}) \\
\frac{d}{dt}\widehat{d}&=& -i[\hat{d},H] =  -i(\omega^{'}_{NV}\hat{d}+Je^{-i\theta}\hat{d}-i\Gamma_{d}\hat{d}) 
\end{eqnarray}

We assume that, after driving the system for a long time, we obtain a
steady state and so we obtain:

\begin{equation}
 (\frac{d}{dt}\hat{c})_{t\rightarrow\infty}=(\frac{d}{dt}\hat{b})_{t\rightarrow\infty}=(\frac{d}{dt}\hat{d})_{t\rightarrow\infty}
  =0
\end{equation}


By solving these, the excited probability of the qubit is described as:
\begin{equation}
\langle\hat{c}^{\dagger}\hat{c}\rangle=(\frac{\lambda}{2})^{2}\times
 |\frac{[(i\Gamma_{b}-\omega^{'}_{b})(i\Gamma_{d}
 -\omega^{'}_{d})-J^{2}]}{(i\Gamma_{c}-\omega^{'}_{c})[(i\Gamma_{b}-\omega^{'}_{b})(i\Gamma_{d}-\omega^{'}_{d})-J^{2}]-g^{2}
 (i\Gamma_{d}-\omega^{'}_{d})}|^{2}
\end{equation}
where $\omega^{'} _{FQ}=\omega_{FQ}-\omega$ and $\omega^{'}
_{NV}=\omega_{NV}-\omega$.
We plot this in Fig. \ref{fig:THOMspectrum}, and this can reproduce three
peaks that are observed in this hybrid system \cite{zhudark2014}.


From the THOM, we confirm that the phase $\theta$ in the expression of
 coupling between the dark state and the bright state would not affect
 the spectroscopic measurement results, and therefore, we set $\theta
 =0$ throughout
 this paper.

 \subsubsection{Properties of the THOM with a resonant condition}
 \begin{figure}[h!]
\par
\centering
\includegraphics[scale=0.51]{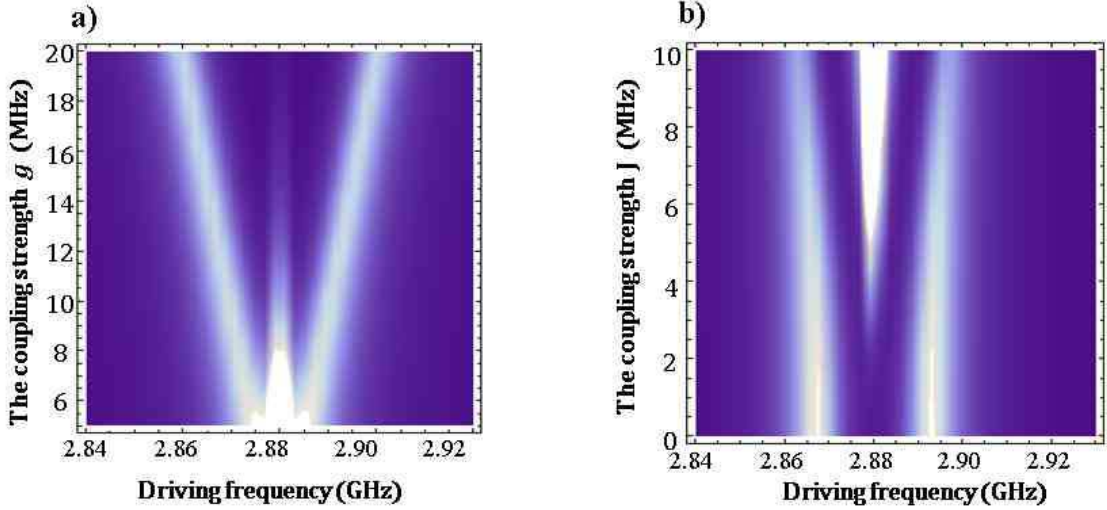}
\par
\caption{Density plot of the spectroscopy of \textcolor{black}{the THOM} to describe the dependency on
  $J$ and $g$
  \textcolor{black}{where} $\Gamma_{FQ}=0.30\times2\pi$ MHz,
  $\Gamma_d=0.50\times2\pi$ MHz, $\Gamma_b=6.40\times2\pi$ MHz,
  and $\lambda=1\times2\pi$ MHz.
  In the picture (a), 
  \textcolor{black}{as the value of $g$ increases}, the middle peak becomes darker.
  Meanwhile, the two side peaks will be largely split for a larger
  $g$. On the other hand, in the picture (b),
  \textcolor{black}{as the value of $J$ increases,
  the effective coupling between the dark state and the flux qubit
  becomes stronger where the bright state mediates the interaction}, which makes the middle peak higher than the other two peaks.}
\label{fig:Jg}
\end{figure}
\begin{figure}[ht!]
\par
\centering
\includegraphics[scale=0.42]{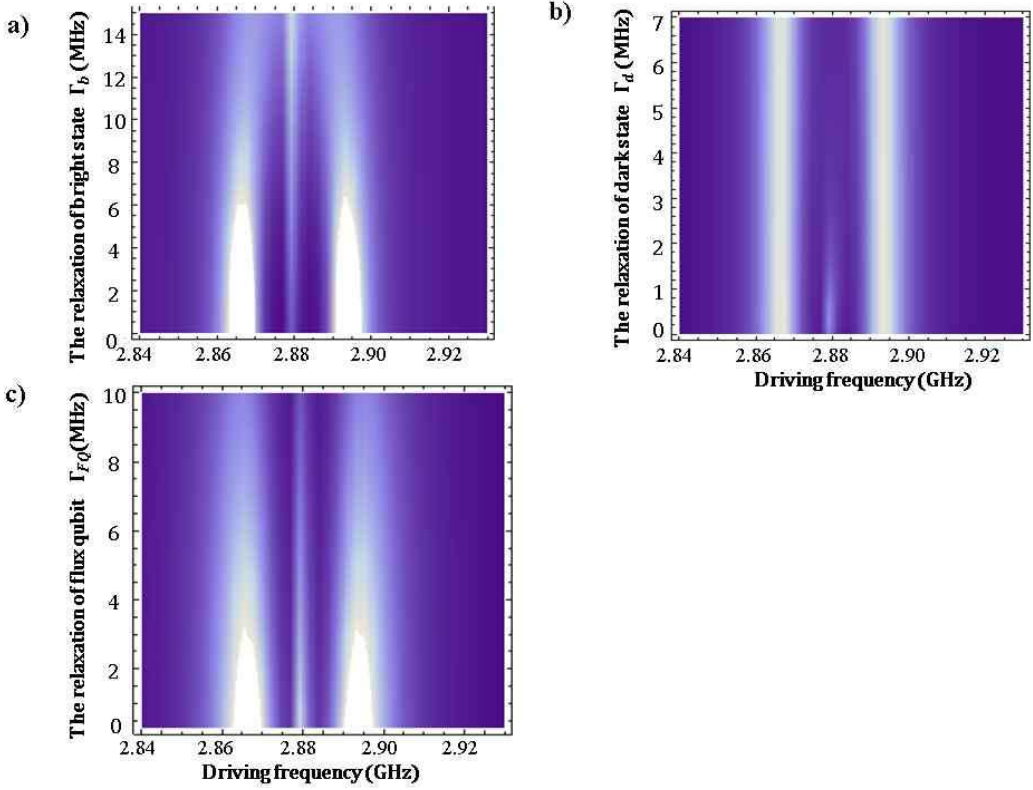}
\par
\caption{Density plot of the spectroscopy to describe the dependency on
 the decay rate $\Gamma_{b}$, $\Gamma_{d}$,
 and $\Gamma_{FQ}$, respectively.
 We fix the other parameters
 $g=13.0\times2\pi$ MHz; $J=3.46\times2\pi$ MHz;
 $\lambda=1\times2\pi$ MHz and then change the $\Gamma_{b}$,
 $\Gamma_{d}$, and $\Gamma_{FQ}$ values.
 These three pictures (a), (b) and (c) correspond to the excited
 population varieties with $\Gamma_{b}$,
 $\Gamma_{d}$, and $\Gamma_{FQ}$. 
 \textcolor{black}{In}
 the picture (a), $\Gamma_{b}$
 mainly affects the width of the side peak which becomes broader as the
 value of $\Gamma_{b}$ increases.
 On the other hand, the width of the middle peak is insensitive against
 $\Gamma_{b}$.
 \textcolor{black}{In} the picture (b),
 the relaxation rate of the dark sate $\Gamma_{d}$ has a significant
 contribution to the excited population
 of middle peak but has a smaller influence on the side peaks. The
 picture (c) shows the effect of $\Gamma_{FQ}$,
 and it change the width of the side peaks and has insignificant effect
 on the middle peak.
 The side peaks decay rapidly as $\Gamma_{FQ}$ increase.}
\label{fig:gamma}
\end{figure}

 We \textcolor{black}{investigate} the behavior of this THOM for several parameters \textcolor{black}{when the
 flux qubit is resonant with the NV$^-$ centers}. From the
simulation result, the hybrid system contains three excited states, and
the middle peak in the spectroscopy denotes the hybrid dark state whose
width is much narrower than that of two side peaks. In order to \textcolor{black}{understand}
the effect of each parameter, we draw a density plot and show the
relationship between the parameters and spectroscopy (shown in
Fig.\ref{fig:Jg} and Fig.\ref{fig:gamma}).

\textcolor{black}{As} the value of $g$ increases,
the separation between two peaks becomes larger (shown in
Fig.\ref{fig:Jg} a).
This can be explained by an increase of the \textcolor{black}{vacuum Rabi
splitting that is determined
by the $J$ and $g$ as follows.}
 \begin{eqnarray}
   \Delta E_{right}-\Delta
  E_{left}
  =2\hbar\sqrt{g^{2}+J^{2}}
 \label{equ:distance}
 \end{eqnarray}
 \textcolor{black}{where} the value of $g$ depends on both a persistent current of the flux
 qubit and the distance between NV$^-$ center and flux qubit.
 The value of  $J$ is determined by the inhomogeneous effect \textcolor{black}{of} NV$^-$ centers.
 Since $g$ is usually much larger than $J$, we have $\Delta
 E_{right}-\Delta E_{left}\simeq 2\hbar(g+\frac{J^{2}}{g})$ (here, $g\gg
 J$)
 and so this energy gap is mainly determined by the value of $g$. 
The area of the peak is determined by the form of the eigenvectors. From
the expression of $\arrowvert E_{middle}\rangle$, as $g$ increases, the
weight of $\arrowvert D\downarrow \rangle$ increases while the weight of
the flux qubit $\arrowvert 0\uparrow \rangle$
 decreases. \textcolor{black}{This means that}, for
a large $g$, it becomes more difficult to detect the dark state by the
spectroscopy where only the flux qubit excitation is measured, and the
middle peak disappears (See the Fig \ref{fig:Jg} a).

 \textcolor{black}{As the value of} $J$ increases, the coupling
between the dark state and bright state becomes stronger, which induces a stronger hybridization between dark state and flux qubit. Since the
weight of $\arrowvert0\uparrow \rangle$  increases with enlarging $J$,
the area of the middle peak becomes larger (See the Fig \ref{fig:Jg} b).

If we change the decay rate $\Gamma_{b}$ of the bright state, it will
make two sides peaks broader (shown in Fig.\ref{fig:gamma} a), due to a
shorter life time. For the effect of $\Gamma_{d}$, it will only
contribute to the middle peak and has almost no influence on the side
peaks (shown in Fig.\ref{fig:gamma} b), which is consistent with the
form of the eigenvector described in Eq.\ref{equ:leftvector} and
Eq.\ref{equ:rightvector} $(J\ll g)$. The flux qubit is directly coupled
with bright state and is coupled with the dark state
indirectly. Therefore, the flux qubit decay rate $\Gamma_{FQ}$  mainly
affects the width of the side peaks, while it has smaller effect on the
middle peak (shown in Fig.\ref{fig:gamma} c).

\subsubsection{Properties of the THOM with a detuning}
\begin{figure}[h!]
\par
\centering
\includegraphics[scale=0.3]{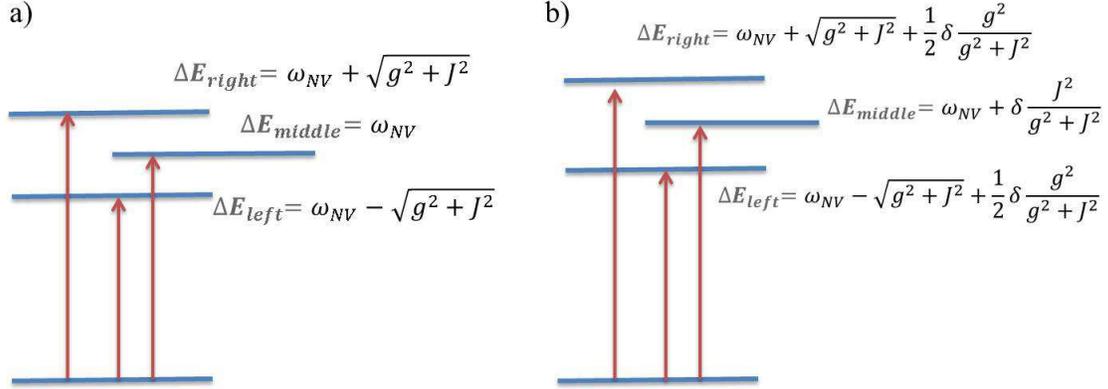}
\par
\caption{The energy level diagram with and without detuning. Picture (a)
 shows
 the energy level and energy gap between ground state and each excited
 state without detuning. Picture
 (b) shows the effect of inducing the detuning between flux qubit and
 NV$^-$ center.
 From the energy level diagram, the frequency of each peak
 will
 cause \textcolor{black}{a} blue shift depending on the $\delta$ value.}
\label{fig:energyleveldetunning}
\end{figure}
Next we consider a case to add the detuning between the flux qubit and
NV$^-$ center. 
\textcolor{black}{We define the detuning} as $\delta =\omega _{FQ}-\omega
_{NV}$. As long as the detuning is much smaller than other
parameters, it is valid to regard the detuning as the perturbation to
the system.  We rewrite the Hamiltonian as following:
\begin{equation}
H=\hbar (\omega _{NV}+\delta)\hat{c}^{\dagger }\hat{c}+\hbar \omega
 _{NV}\hat{b}^{\dagger}\hat{b} +
 \hbar \omega _{NV}\hat{d}^{\dagger}\hat{d} +\hbar g
 (\hat{c}^{\dagger}\hat{b}+\hat{c}\hat{b}^{\dagger })+
 \hbar J(\hat{b}^{\dagger}\hat{d}+\hat{b}\hat{d}^{\dagger})
\label{equ:hamiltonian}
 \end{equation}

 \begin{figure}[ht!]
\par
\centering
\includegraphics[scale=0.4]{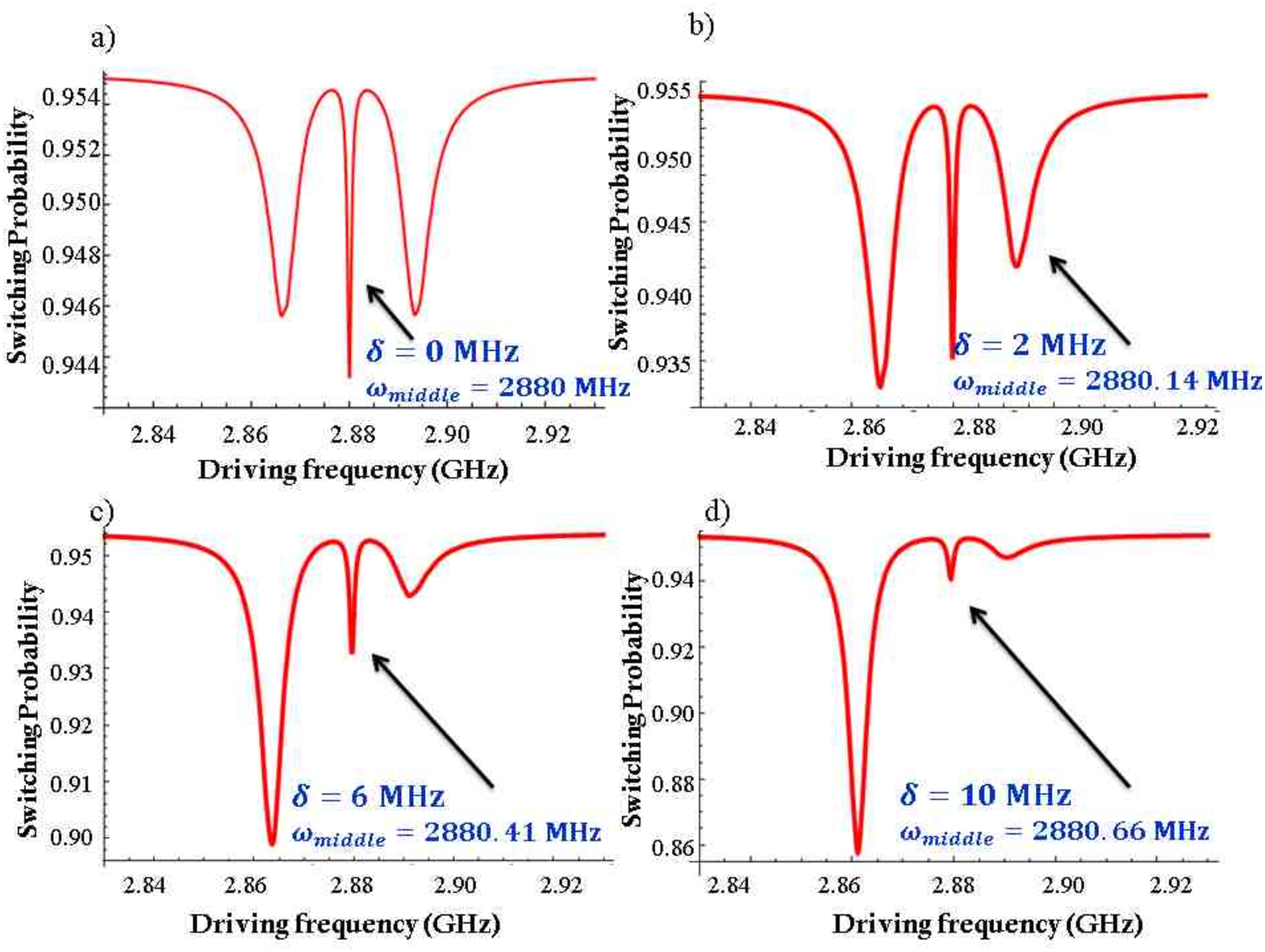}
\par
\caption{Energy shift of the middle peak against detuning
  \textcolor{black}{for MHOM}. Each picture presents for the case of the
  detuning as 0, 2 MHz, 6 MHz and 10 MHz, respectively.
  The detuning makes it difficult to excite the left and middle peak
  whose frequency is far from the energy of the flux qubit.
  Here we choose the same parameters as in Fig \ref{fig:energy spectrum
  MHOM} except of the frequency of the flux qubit which
  is equal to the sum of the detuning and frequency of NV$^-$ center.}
\label{fig:detning}
\end{figure}
\begin{figure}[ht!]
\par
\centering
\includegraphics[scale=0.4]{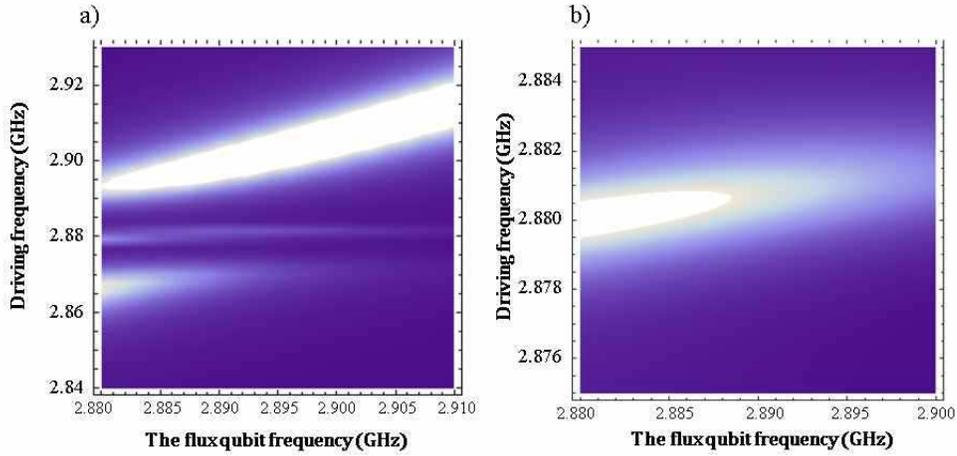}
\par
\caption{Numerical simulation by solving THOM where the x axis denotes
 the flux-qubit frequency and y axis denotes the microwave frequency.
 (a) Density plot of the spectroscopy. (b) Magnified view of the
 spectroscopy around 2.88 GHz.
 These pictures clearly show that every peak has a finite energy shift
 if we add the detuning.
 Here, we use the same parameters as in Fig. \ref{fig:THOMspectrum}.}
\label{fig:3Ddetuning}
\end{figure}
 By diagonalising the Hamiltonian and using the perturbation theory, we
can calculate the eigenvalues \textcolor{black}{and eigenvectors}
(shown in
Fig.\ref{fig:energyleveldetunning}).
\begin{equation}
 \Delta E_{left}=E_{left}- E_{0}\approx\hbar(\omega_{NV}-\sqrt{g^{2}+J^{2}}+\frac{1}{2}\delta\frac{g^{2}}{g^{2}+J^{2}})
 \label{equ:left}
\end{equation}
\begin{eqnarray}
\arrowvert E_{left}\rangle=
 -\frac{1}{\sqrt{2}}[1+\frac{g^2\delta}{4(g^2+J^2)^{3/2}}]\cdot\arrowvert
 B\downarrow\rangle\nonumber \\
 +\frac{1}{\sqrt{2}}\frac{g}{\sqrt{g^2+J^2}}[1-\frac{g^2\delta}{4(g^2+J^2)^{3/2}}-\frac{J^2\delta}{(g^2+J^2)^{3/2}}]\cdot\arrowvert
  0\uparrow\rangle\nonumber  \\
+\frac{1}{\sqrt{2}}\frac{J}{\sqrt{g^2+J^2}}[1-\frac{g^2\delta}{4(g^2+J^2)^{3/2}}+\frac{J^2\delta}{(g^2+J^2)^{3/2}}]\cdot\arrowvert D\downarrow\rangle
\end{eqnarray}
\begin{equation}
 \Delta E_{middle}=E_{middle}- E_{0}\approx\hbar(\omega_{NV}+\delta\frac{J^{2}}{g^{2}+J^{2}})
 \label{equ:middle}
\end{equation}
\begin{equation}
\arrowvert E_{middle}\rangle= -\frac{J}{\sqrt{g^2+J^2}}\arrowvert
 0\uparrow\rangle+\frac{g}{\sqrt{g^2+J^2}}\arrowvert D\downarrow\rangle
 +\delta\frac{gJ}{(g^2+J^2)^{3/2}}\arrowvert B\downarrow\rangle
\end{equation}
\begin{equation}
 \Delta E_{right}=E_{right}- E_{0}\approx\hbar(\omega_{NV}+\sqrt{g^{2}+J^{2}}+\frac{1}{2}\delta\frac{g^{2}}{g^{2}+J^{2}})
 \label{equ:right}
\end{equation}
\begin{eqnarray}
\arrowvert E_{right}\rangle=
 -\frac{1}{\sqrt{2}}[1-\frac{g^2\delta}{4(g^2+J^2)^{3/2}}]\cdot\arrowvert
 B\downarrow\rangle\nonumber \\
 +\frac{1}{\sqrt{2}}\frac{g}{\sqrt{g^2+J^2}}[1+\frac{g^2\delta}{4(g^2+J^2)^{3/2}}+\frac{J^2\delta}{(g^2+J^2)^{3/2}}]\cdot\arrowvert
  0\uparrow\rangle\nonumber  \\
+\frac{1}{\sqrt{2}}\frac{J}{\sqrt{g^2+J^2}}[1+\frac{g^2\delta}{4(g^2+J^2)^{3/2}}-\frac{J^2\delta}{(g^2+J^2)^{3/2}}]\cdot\arrowvert D\downarrow\rangle
\end{eqnarray}
\textcolor{black}{These results show that, as}
the detuning increases, the energy difference between the excited states
and the ground state would increase. The amount of change depends on
$g$, $J$, and $\delta$ (shown in Fig.\ref{fig:detning}).
\textcolor{black}{It is worth mentioning that the} middle peaks has a smaller energy shift
than that of the other two side peaks for $g\gg J$.

From the spectroscopy calculated by MHOM, we have also confirmed the
fact that the frequency of middle peak can be shifted by the detuning
(See Fig.\ref{fig:detning} and Fig.\ref{fig:3Ddetuning}). Also, \textcolor{black}{the left
and right
peaks have} an energy shift for the detuning, and the energy shift is much
larger than that of the middle peak. ( See the Fig.\ref{fig:detning} and
Fig.\ref{fig:3Ddetuning})

\end{document}